# NEW TECHNIQUE FOR PROPOSING NETWORK'S TOPOLOGY USING GPS AND GIS

Ayad Ghany Ismaeel

Department of Information Systems Engineering- Erbil Technical College- Technical Education Foundation, Erbil-Iraq
`dr_a_gh_i@yahoo.com`

## ABSTRACT

*The problem of proposed topology for network comes when using Prim's algorithm with default distance (unrealistic distances) between network's nodes and don't care about the lakes, high hills, buildings, etc. This problem will cause incorrect estimations for cost (budget) of requirements like the media (optic fibre) and the number or type of Access-points, regenerator, Optic Amplifier, etc.*

*This paper proposed a new technique of implementing Prim's algorithm to obtain realistic topology using realistic distances between network's nodes via Global* Positioning *System GPS and Geographic Information Systems GIS packages. Applying the new technique on academic institutes network of Erbil city from view of media (optic fibre) shows that there is disability in cost (budget) of the media which is needed ≅4 times if implement default Prim's algorithm (don't using GPS & GIS) base on unrealistic distances between the nodes.*

## 1. INTRODUCTION

The important stage for proposed any modern (optic fiber) Internet Protocol **IP** networks (*Internet, Intranet, or Extranet*) which are common in use now in the world is the infrastructure, this infrastructure will contain Access-points (**Routers/** Asynchronous Transfer Mode **ATM**s), Synchronous Optic Network **SONET**, the backbone network which is constructed from Dense Wavelength Division Multiplexing **DWDM**s [3, 10].

The connection of the access points, SONET, or backbone's nodes needs proposed topology (*the way which are connect the nodes of network*), there are multiple modeling method [8], algorithms, etc use to obtain the topology, the important one and commonly in use is Prim's algorithm which gives an efficient (*low cost and high traffic*) topology because it is important application of *Graph Theory* base on Minimum Spanning Tree **MST** as show following in the Prim's algorithm [1]:

**Input:** *A weighted connected graph G with n vertices and m edges*
**Output:** *A Minimum Spanning Tree T for G*

*Q = new heap-based priority queue*
*s = a vertex of G {pick up any vertex s of G}*
*Initialize T to null*
*for all v ∈ G.vertices()*
*if (v = s) then setDistance(v, 0)          {set the key to zero}*
*else setDistance(v, ∞)*
*setParent(v,Ø)                          {parent edge of each vertex is null}*
*Initialize the Q with an item ((u, null), D[u] for each vertex u, where (u, null)*
*is the element and D[u] is the key. D[u] is the distance of u*
*while ¬ Q.isEmpty()*
*(u, e) ← Q.removeMin()Add vertex u and edge e to T*
*for all e ∈  G.incidentEdges(u)*
*z ← G.opposite(u,e)      {for each vertex z adjacent to u such that z is in Q}*
*r ← weight(e)              {= w(u, z) }*
*if r < getDistance(z)*
*setDistance(z, r)              {update the D[z ] in Q}*
*setParent(z, e)                {update the parent of z in Q}*
*return the tree T*

## 2. RELATED WORK

The problem of suggestion topology comes when using Prim's algorithm with distances between the nodes of network taken by default, i.e. depending on distance of maps (e.g. the distance between countries, cities, villages) or using unrealistic distances, it means don't care about natural preventives (the lakes/rivers, high hills, buildings, etc) so become the suggested topology in common unrealistic.

This problem commonly will cause:

i. Incorrect estimations of the needs of the expensive media which is used in modern IP network (*e.g. optic fiber which is fast and unlimited bandwidth*) and mistake in numbers of other requirements which are needed in IP network like Access-points, SONET, DWDM, regenerator, OA, etc.

ii. Shortage in budgets because the designers make their computations base on unrealistic distances, and that will cause the implement of proposed IP network later impossible, i.e. the designers haven't enough devices, media, equipments, etc for implement.

iii. Later can't done the maintenance for IP Network in easy way (low costs) because may have natural preventives.

## 3. THE AIM OF RESEARCH

Solving the problem of unrealistic distance between the nodes of IP network and take in consideration all natural preventives, that will give realistic topology, i.e. will obtain real costs to implement and later doing an easy maintenance of networks.

The main goal of this research is obtaining realistic topology using Prim's algorithm base on real distance between nodes rather than default or unrealistic distances (don't care about real distance and natural preventives), so will need a new technique base on **GPS, UTM**, and **GIS** packages which are explaining later.

## 4. NEW TECHNIQUE OF PROPOSING NETWORK'S TOPOLOGY

The new technique of suggestion a network's topology limited in a flowchart involve **6** steps as shown in Figure 1:

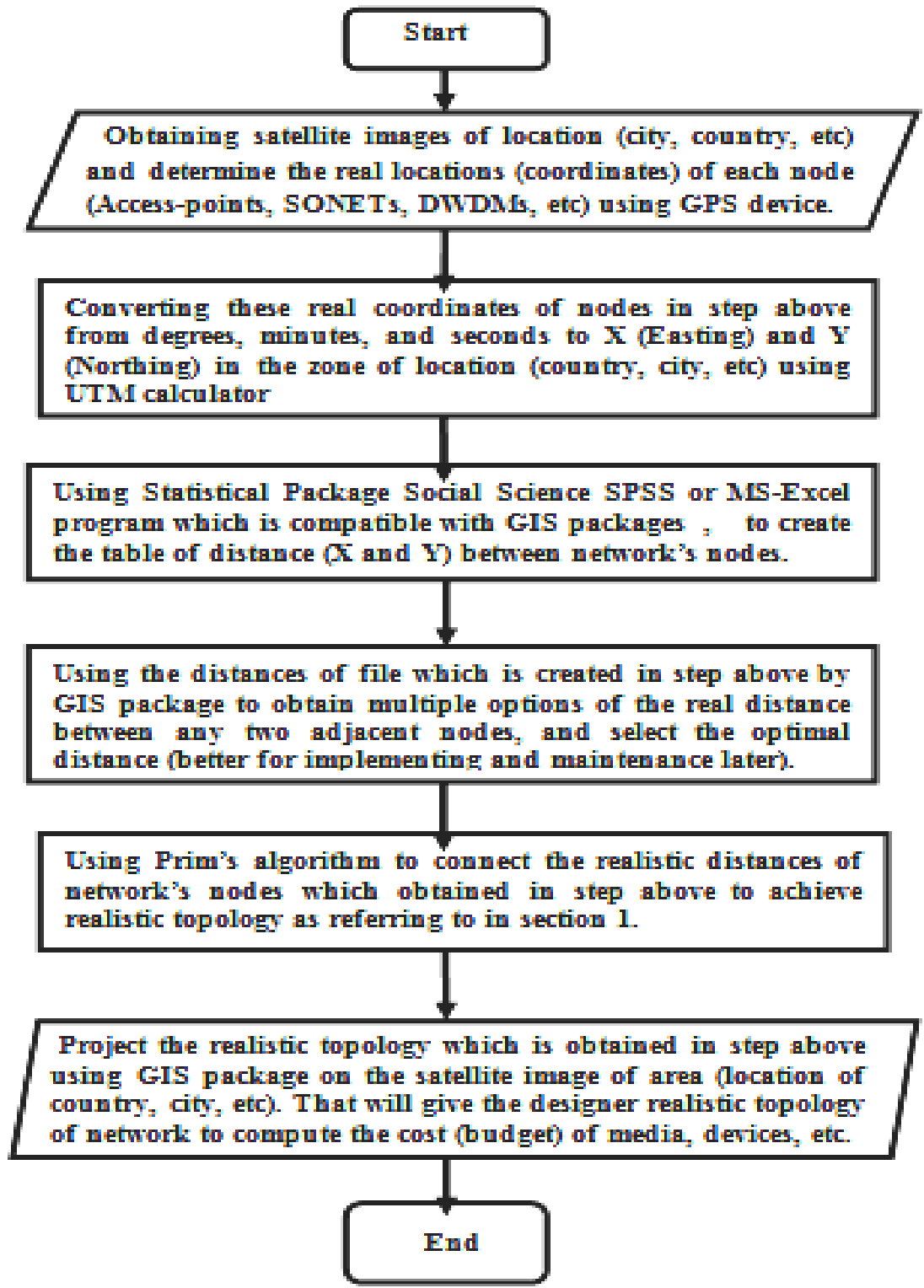

Figure 1. Flowchart of new technique for proposing network's topology

# 5. IMPLEMENTING THE NEW TECHNIQUE

Applying the new technique of proposing network's topology on the campuses of **Academic** Institutes **Net**work in Erbil city is called **AcademicNet** to obtain realistic topology. This application will determine only the media (fibres) need using a new technique as follow:

**A.** Collecting information about the AcadmeicNet's campuses in Kurdistan-region from **M**inistry **o**f **H**igher **E**ducation and Scientific Research **MOHE** as shown in Table 1, then get satellite image of Erbil city from Erbil's government.

Table 1: Reveals the AcademicNet's campuses.

| Campuses Network | Number |
|---|---|
| University's Colleges + Ministry | 24 |
| Institutions | 3 |
| Private University | 2 |
| Total | 29 |

**B.** Find the coordinates of each campus of Table 1 by visiting each campus using GPS, then convert these coordinates to Easting (X) and Northing (Y) using UTM calculator [2,5], which is shown in Figure 2.

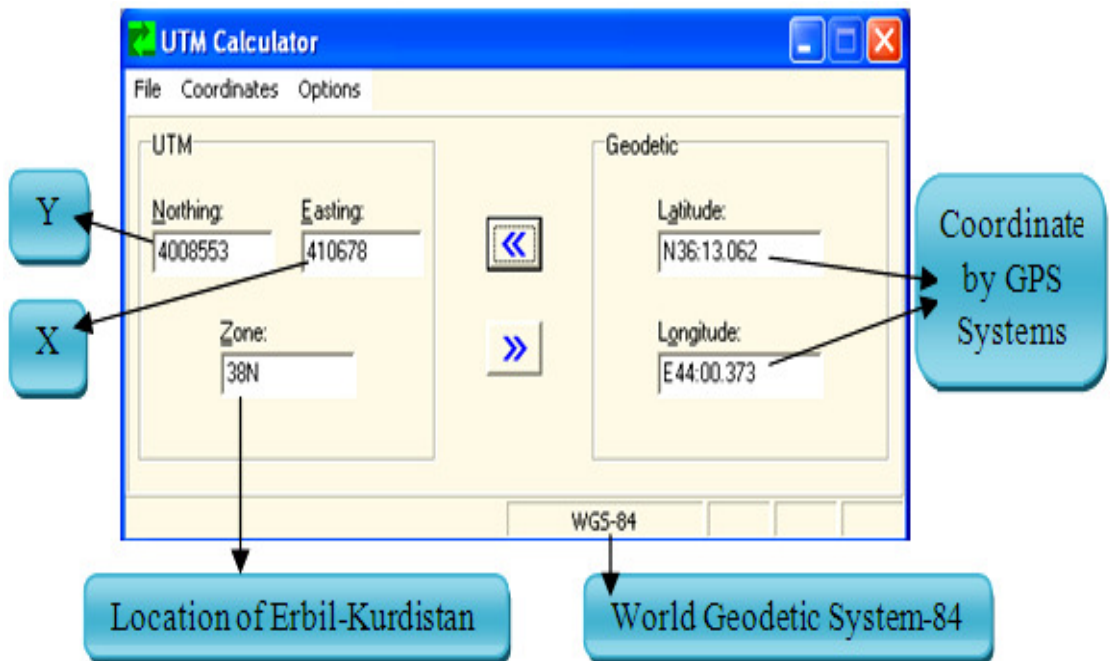

Figure 2: UTM Calculator Process

The converting of each campus reveals in Table 2 [6].

Table 2: Converts GPS's coordinates to X & Y by UTM calculator

**C.** Create Table 2 using MS-Excel, to become ready to use by ArcGis 9.2 as GIS package [9, 4], as referring to in flowchart in section 3

| No. | Campus of AcademicNet's network | GPS Coordinate | | UTM Calculator | |
|---|---|---|---|---|---|
| | | Latitude | Longitude | Northing Y | Easting X |
| 1 | Ishek university | N36:11:38.90 | E44:01:29.20 | 4005923 | 412321 |
| 2 | Shahana Britches collage | N36:13:03.40 | E44:00:33.80 | 4008540 | 410963 |
| 3 | SABIS university/Education collage | N36:12:27.40 | E44:00:12.70 | 4007436 | 410424 |
| 4 | Dijla collage | N36:13:06.10 | E44:00:37.30 | 4008621 | 411051 |
| 5 | Islamic Science collage | N36:11:13.20 | E44:01:03.90 | 4005137 | 411680 |
| 6 | Dentist collage | N36:11:58.50 | E44:01:15.00 | 4006530 | 411972 |
| 7 | Medicine collage | N36:11:52.50 | E44:01:10.90 | 4006346 | 411868 |
| 8 | Fine art collage | N36:11:05.30 | E44:00:10.90 | 4004907 | 410354 |
| 9 | Kurdistan university | N36:10:53.90 | E44:00:20.60 | 4004552 | 410592 |
| 10 | Medicine university | N36:11:24.40 | E44:02:23.80 | 4005463 | 413680 |
| 11 | Management &Economic collage | N36:09:49.50 | E44:00:57.20 | 4002559 | 411486 |
| 12 | Dean of agriculture collage | N36:09:38.90 | E44:00:52.60 | 4002234 | 411368 |
| 13 | Art collage | N36:09:34.20 | E44:00:50.30 | 4002091 | 411309 |
| 14 | MOHE | N36:09:39.30 | E44:00:53.70 | 4002245 | 411397 |
| 15 | Human education | N36:09:43.10 | E44:00:55.00 | 4002364 | 411430 |
| 16 | Fine art institute | N36:09:29.60 | E44:00:45.70 | 4001950 | 411194 |
| 17 | Hawler medicine institute | N36:09:31.30 | E44:00:47.30 | 4002000 | 411237 |
| 18 | Agriculture collage | N36:09:44.50 | E44:00:49.10 | 4002409 | 411283 |
| 19 | Science collage | N36:09:10.90 | E44:01:16.50 | 4001364 | 411957 |
| 20 | presidency of Salahalddin university | N36:09:50.50 | E44:00:58.40 | 4002590 | 411516 |
| 21 | Education collage | N36:09:44.40 | E44:01:00.00 | 4002401 | 411554 |
| 22 | Teaching collage | N36:08:56.40 | E44:01:09.70 | 4000920 | 411782 |
| 23 | Hawler Technical institute | N36:08:37.70 | E44:01:01.50 | 4000374 | 411571 |
| 24 | Law collage | N36:08:35.70 | E44:01:40.30 | 4000274 | 412541 |
| 25 | Collage of sports education | N36:08:36.20 | E44:01:36.00 | 4000291 | 412434 |
| 26 | Collage of Engineering | N36:08:33.10 | E44:01:25.40 | 4000198 | 412167 |
| 27 | Hawler Technical collage | N36:08:36.40 | E44:02:17.20 | 4000288 | 413464 |
| 28 | Hawler computer institute | N36:10:03.40 | E43:58:22.90 | 4003028 | 407637 |
| 29 | Gihan university | N36:10:16.10 | E43:57:52.00 | 4003426 | 406867 |

**D.** Project the coordinates of each campus (X and Y) from Excel file which created in C on satellite image of Erbil city using ArcGis 9.2, then select the optimal distance from multiple options between each adjacent campuses (lowest cost and easy maintenance in future) of AcademicNet as shown in Table 3.

Table3: Optimal/Realistic distances in meters between adjacent campuses **(via ArcGis 9.2)**

| No | Campus-to-Campus (Adjacent Two) | Optimal Distance | No | Campus-to-Campus (Adjacent Two) | Optimal Distance |
|----|--------------------------------|-----|----|--------------------------------|-----|
| 1 | MOHE-to-Hawler Computer Institute | 4740 | 27 | Fine art institute-to-Agriculture  College | 600 |
| 2 | MOHE-to-Technical Institute | 2360 | 28 | Fine art institute-to-Human education | 760 |
| 3 | MOHE-to-Art College | 170 | 29 | Fine art institute-to-Economic | 940 |
| 4 | MOHE-to-Agriculture College | 170 | 30 | Fine art institute-to-Medicine university | 5850 |
| 5 | Hawler Computer Institute-to-Science | 6100 | 31 | Slahaldin presidency-to-Agriculture deans | 515 |
| 6 | Hawler Computer Institute-to-teaching College | 6260 | 32 | Slahaldin presidency-to-Human education | 225 |
| 7 | Hawler Computer Institute-to-sports | 7740 | 33 | Slahaldin presidency-to-Economic | 70 |
| 8 | Hawler Computer Institute-to-Economic | 4890 | 34 | Agriculture deans-to-Humane education | 270 |
| 9 | Technical Institute-to-Art College | 1750 | 35 | Agriculture deans-to-Economic | 470 |
| 10 | Technical Institute-to-Law College | 1400 | 36 | Human education-to-Economic | 200 |
| 11 | Technical Institute-to-Technical College | 2520 | 37 | Kurdistan university -to-Islamic science | 1700 |
| 12 | Art College-to-Education science | 315 | 38 | medicine institute-to-Kurdistan university | 3030 |
| 13 | Art College-to-Agriculture College | 435 | 39 | Kurdistan university-to-Fine arts collage | 680 |
| 14 | Science College-to-Teaching College | 680 | 40 | Kurdistan university-to-Economic | 2610 |
| 15 | Science College-to-Sports | 2150 | 41 | Kurdistan university-to-Shahana British | 4620 |
| 16 | Science College-to-Eng College | 1700 | 42 | Fine arts collage-to-medicine university | 4540 |
| 17 | Science College-to-Dijla Private University | 8420 | 43 | Fine arts collage-to-Fine art institute | 3500 |
| 18 | Teaching College-to-Sports  College | 1030 | 44 | Islamic science-to-Medicine institute | 3760 |
| 19 | Teaching College-to-Eng College | 1000 | 45 | Islamic science-to-Ishek university | 1680 |
| 20 | Sports  College-to-Eng College | 1500 | 46 | Collage of medicine-to-Ishek university | 1030 |
| 21 | Law  College-to-Agriculture College | 3200 | 47 | Collage of medicine-to-Shahana British | 3180 |
| 22 | Law  College-to-Technical College | 2445 | 48 | Shahana British-to-Sabis | 2135 |
| 23 | Education science College-to-Agriculture College | 340 | 49 | Sabis-to-Dijla | 2050 |
| 24 | Education science  College-to-Technical College | 4450 | 50 | Dijla-to-Gihan | 8100 |
| 25 | Agriculture College-to-Law  College | 2850 | 51 | Collage of medicine-to-Dentist | 560 |
| 26 | Fine art institute-to-Salahaldin presidency | 990 | | | |

**E.** Use the 51 distance of Table 3 in implementing Prim's algorithm via java applet to obtain realistic topology of AcademicNet which contains 29 campuses (nodes) as shown in Figure 3.

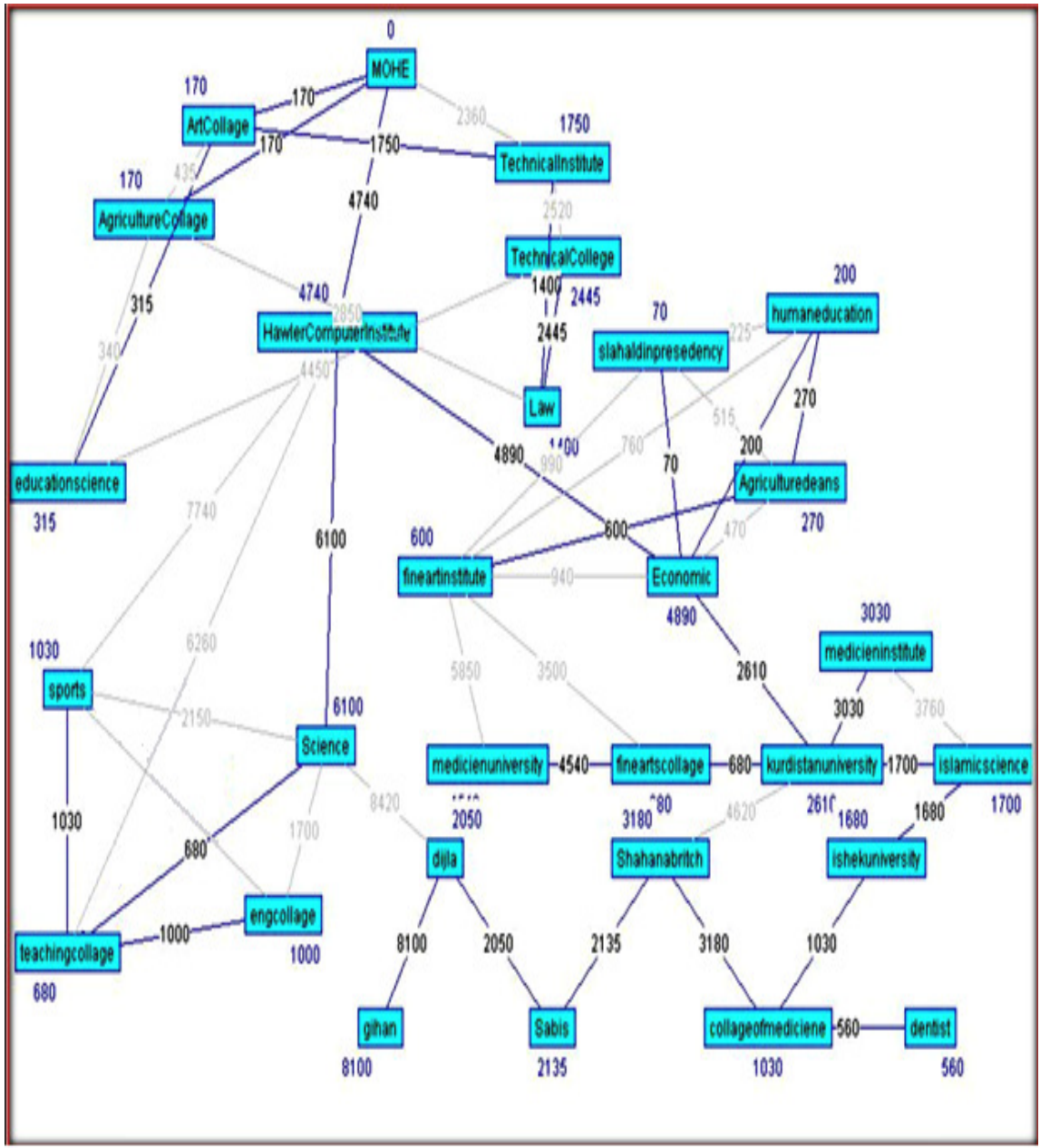

Figure 3: Implement Prim's algorithm for AcademicNet's topology using java applet

**F.** Project the realistic topology of AcademicNet which is obtained in E on the satellite image of Erbil city via ArcGis 9.2, as shown in Figure 4 to make AcademicNet ready for designer to obtain accurate computation for implementation and maintenance in future.

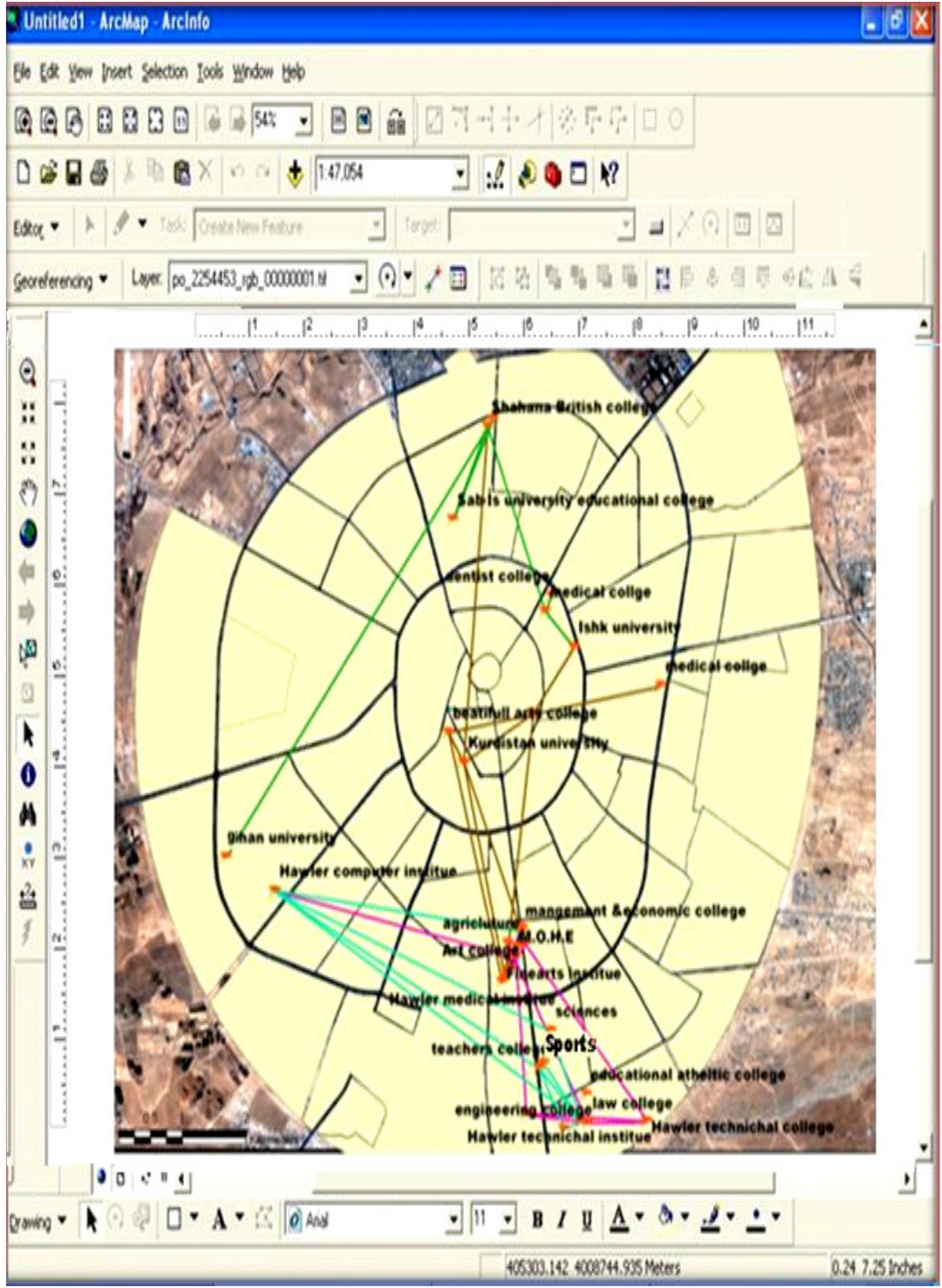

Figure 4: Projection a realistic AcademicNet's topology (obtained in E) using ArcGis 9.2

### 5.1 Comparing the topology of new technique with default topology

Table 4 shows comparison between the new technique of suggestion topology of network which is applied on AcademicNet with another topology of the same campuses network for Erbil city within "Designing of Kurdistan Scientific Information Intranet Network" is called KSIIN, the topology of campuses in KSIIN is obtained by implementing Prim's algorithm base on default distances (i.e. don't using GPS & GIS) between the campuses [7].

Table 4: Comparison the topology of new technique with the default topology

| No. | Characteristic | AcademicNet's Topology using new technique of GPS & GIS | Topology of campuses in KSIIN which obtained without using GPS & GIS |
|---|---|---|---|
| 1 | No. of campuses & edges | 29 campus and 28 edge | 26 campus and 25 edge |
| 2 | Coordinate & distance | Depended on GPS device and real distance using GIS program | Depended on pixels of monitor and default (unreal) distance |
| 3 | Software | Additional of Java applet will use GPS, UTM, Excel/SPSS, and GIS program. | Using java applet program only |
| 4 | Environment of Projection | The projections of nodes are applied on satellite image of Erbil city using GIS | The projections of nodes apply on the map of Erbil city |
| 5 | Length of Media(Fiber) | **57125** meter | **11465** meter |

### 5.2 The experimental results

The new technique of proposing topology is fault-tolerance in determination of the distances or coordinates for campuses, this cause a different topologies for the following reasons:

**a.** Using the GPS and there aren't previously fixed points (labels) for campuses that make at another time (e.g. implementation stage) or when another person determine the coordinates of nodes (campuses) this will cause by differing distances.

**b.** Depending on un-recent satellite image for the location (city, country, etc) that will make later the projection of coordinates using GIS with recent satellite image gives differing distances.

For example Table 5 reveals two different determinations of distances for the same collection of campuses in AcademicNet, the difference just in (**1**) and (**2**) of Table 5 supposing happened for the reasons (**a**) or (**b**) above that difference in distances will give different topologies. Figure 5 appears the topology for the collection of AcademicNet's nodes using the first determination of distances in Table 5 with the shortest path (minimum distance which are colour by blue in Figure 5) for the topology calculated (**24000** meter), while Figure 6 reveals another topology for the second determination of distances in Table 5 for the same collection of AcademicNet's nodes, by shortest path calculated (**22120** meter).

Table 5: Different determination of distances for collection of AcademicNet's nodes

| No | Campus-to-Campus(Adjacent Two) | First Determination in Meter | Second Determination in Meter |
|----|-------------------------------|------------------------------|-------------------------------|
| **1** | **Hawler Computer Institute -to-Science** | **6200** | **6100** |
| **2** | **Hawler Computer Institute -to-teaching College** | **6160** | **6260** |
| 3 | Hawler Computer Institute -to-sports | 7740 | 7740 |
| 4 | Hawler Computer Institute -to-Economic | 4890 | 4890 |
| 5 | Science -to-teaching College | 680 | 680 |
| 6 | Science -to-sports | 2150 | 2150 |
| 7 | Science -to-engcollege | 1700 | 1700 |
| 8 | Science -to-dijla | 8420 | 8420 |
| 9 | sports -to-engcollege | 1500 | 1500 |
| 10 | teaching College -to-sports | 1030 | 1030 |
| 11 | teaching College -to-engcollege | 1000 | 1000 |

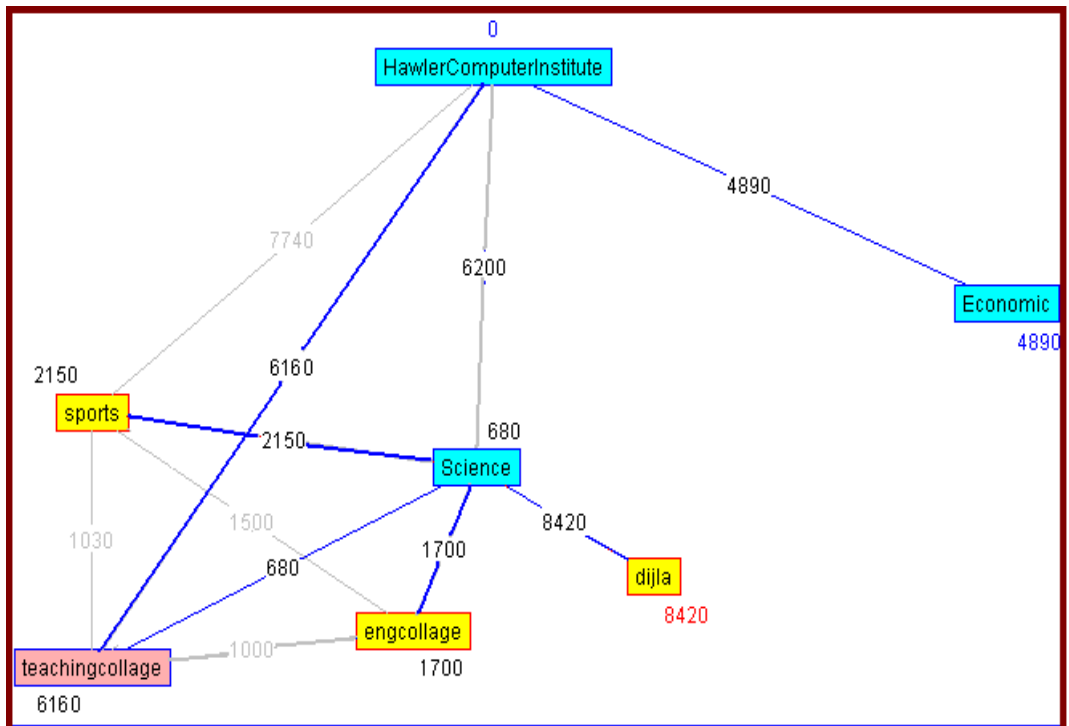

Figure 5: Shows topology of *1st* determination of distances for AcademicNet's nodes collection

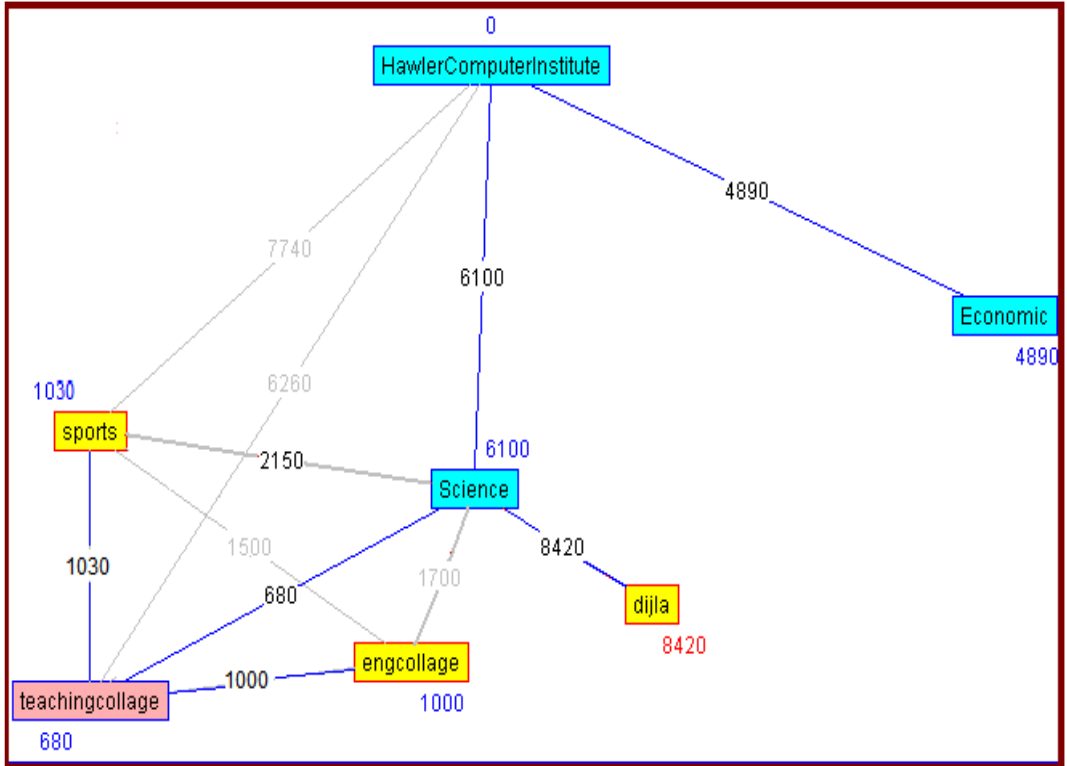

Figure 6: Shows topology of *2nd* determination of distances for AcademicNet's nodes collection

## 6. CONCLUSIONS

Applying the new technique of suggestion topology for network gives the following conclusions:

1. The optic fibers which is needed in AcademicNet's topology using the new technique is different by **45660** meter than topology's campuses in KSIIN which obtained without using GPS and GIS as shown in Table 4; 5, i.e. if take in consideration the extend in academic institutes of Erbil at the interval between proposed the two topologies as referring to in subsection 5.1, will see topology's campuses in KSIIN's computed **20%** only from media, so there is disability in cost (budget) ≅ **4 times** than AcademicNet's topology.

2. The other requirements will affect by that shortage in distances, i.e. the effect can see in following levels:

   i. Network of access-points (campuses), may need ATM switches (which are covered more distance but expensive) more than routers.

   ii. SONET's network the shortage in this level will require more numbers of regenerators, Add Drop Multiplexer **ADM**, etc.

   iii. Backbone (DWDM) network the shortage of distance will cause needing more OAs, Optic Add Drop Multiplexer OADM, etc.

3. Realistic distances of campuses are taking base on streets/roads of Erbil city to avoid the natural preventives this will cause lowest cost for network maintenance in future.

**4.** Avoiding the fault in determination of the distances or coordinates of the network's nodes as referring to in subsection 5.2, by determining previously fixed point (label) for each node and dependence on recent satellite image.

**Author**

**Ayad Ghany Ismaeel** is currently an assistant professor of computer science at department of Information Systems Engineering at Erbil Technical College-Iraq. He received his Ph.D. degree in computer science from University of Technology at Baghdad- Iraq in 2006. His research interest's mobile and IP networks, Web application, GPS, GIS techniques, distributed systems and distributed databases. He is lecturer in postgraduate of few universities in MSc and PhD courses in computer science and software engineering from 2007 till now in Kurdistan-Iraq.

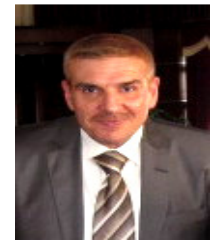